\documentclass[amsmath,amssymb,aps,twocolumn,notitlepage,showpacs,pre,floatfix]{revtex4-1}
\usepackage{graphicx}
\usepackage{bm}
\usepackage{hyperref}
\usepackage{color}
\usepackage[T1]{fontenc}
\usepackage{soul}

\begin{document}

\title{Stationary ordered non-equilibrium states of long-range interacting systems}

\author{Michael Joyce$^1$, Jules Morand$^{2}$ 
and Pascal Viot$^3$ }
\affiliation{$^1$ Laboratoire de Physique  Nucl\'eaire et de Hautes \'Energies, UPMC IN2P3 CNRS 
UMR 7585, Sorbonne Universit\'es, 4, place Jussieu, 75252 Paris Cedex 05, France}
\affiliation{$^2$  
National Institute for Theoretical Physics (NITheP),Stellenbosch 7600, South Africa \& 
Institute of Theoretical Physics, Department of Physics, Stellenbosch University, Stellenbosch 7600, South Africa }
\affiliation{$^3$ Laboratoire de Physique
  Th\'eorique de la Mati\`ere Condens\'ee, UPMC, CNRS  UMR 7600, Sorbonne Universit\'es, 4, place Jussieu, 75252 Paris Cedex 05, France}

\begin{abstract}
Long-range interacting Hamiltonian systems are believed to relax
generically towards non-equilibrium states called ``quasi-stationary" 
because they evolve towards thermodynamic equilibrium very slowly,
on a time-scale diverging with particle number. We show here that, by 
applying a suitable perturbation operator for a finite time interval, we obtain,  
in a family of long-range systems, non-equilibrium states which appear 
to be strictly stationary.  They exist even in the case of a harmonic potential, 
and are characterised by an ordered microscopic phase space structure. We 
give some simple heuristic arguments which predict reasonably 
well some properties of these states.
\end{abstract}
\date{\today}
\maketitle

Long-range interacting systems are ubiquitous in nature and concern diverse physical 
situations (for an overview see e.g. \cite{Campa2014}) from very large length scales 
as self-gravitating bodies in astrophysics \cite{Padmanabhan1990,Binney2008}, to 
very small length scales such as cold atom systems \cite{Chalony2013,Barre2014}, 
plasmas\cite{Nicholson1983},  free-electron lasers\cite{Barre2004} as well as chemotaxis 
in biological systems\cite{Keller1971,Sire2002}.
One of the apparently universal features of these systems is that the dynamics of 
their relaxation to microcanonical equilibrium is extremely slow: the time-scale for 
it is typically found to diverge as a positive power of particle number
(see e.g. \cite{Farouki1982,Yamaguchi2004,Joyce2010,Gabrielli2010a,Teles2010,2017arXiv170101865M}).
On mean-field time-scales, however, the
systems are observed to attain macroscopically stationary
out-of-equilibrium states,  interpreted as stationary states of the Vlasov equation.
These states, known  as ``quasi-stationary'' states (QSS) because of
their slow evolution towards equilibrium, 
are of fundamental physical importance, as they represent essentially 
macroscopic equilibria of a large class of Hamiltonian systems. Many
basic questions about them remain, concerning notably both the mean-field
dynamics leading to the QSS and their long-time relaxation dynamics.
Several authors 
(see e.g. \cite{PhysRevLett.105.040602, Nardini2012, Patelli2012,Chavanis2011}) and we ourselves \cite{Joyce2014,Joyce2016} 
have investigated the interesting issue of whether QSS may or may not persist in the presence of perturbations
to the Hamiltonian dynamics. Our study 
has revealed a rather unexpected and intriguing result which 
is the subject of this Letter: a particular class of perturbations, applied to such systems for a finite time, can drive them
efficiently to states of the $N$ body system which, like QSS, are macroscopically stationary on mean
field times scales but, unlike QSS, do not evolve at all (on the very longest time-scales we can
access numerically). Microscopically these states are very different to typical QSS, 
as they are characterised by a highly regular ``ordered'' structure in phase space corresponding
to periodic or quasi-periodic solutions of the $N$-body Hamiltonian dynamics. 

Specifically we consider here one dimensional long-range interacting systems with 
interparticle potential $V(r)= gr^\alpha$ where $r$ is the distance between particles
and $sgn (g)= sgn (\alpha)$ (i.e. attractive forces). The numerical
results reported below are for the case $\alpha=1$, corresponding to the so-called
``sheet model" , equivalent to  infinite, infinitely thin,  parallel sheets 
moving in three dimensions under Newtonian self-gravity 
(see e.g. \cite{Miller1996,Yawn1997,Joyce2010}  and references therein). 
As detailed in the Supplemental Material, we have explored a large range of 
values for the exponent $\alpha$ in the range $[-1, 2]$ and found our essential
results to hold in this entire range, provided a suitable regularisation of
the singularity at $r=0$ is employed for $\alpha <1$. 
The model with $\alpha=2$ is equivalent to a set of uncoupled  identical 
harmonic oscillators, for which  
the dynamics is thus evidently non-ergodic,  
while in other cases studies suggest that it is expected to be 
ergodic \cite{Milanovic2006,Milanovifmmodecuteclseci1998}, at least 
for a sufficient number of particles. Specifically, for the case $\alpha=1$,
using an analysis of the stability of periodic orbits \cite{Reidl1993}
have concluded that ergodicity applies for $N > 11$. 
Ergodicity breaking is quite ubiquitous for  long range interacting 
systems \cite{Borgonovi2004,Bouchet2008,PhysRevLett.95.240604,Teles2010,Benetti2012}, 
but that we discuss here has a completely different origin to
previous studies. 

Using molecular dynamics simulations, we  study these models with a simple
modification: for a certain finite duration we modify the Hamiltonian dynamics
by implementing a non-trivial collision rule when particles cross one another. 
Assuming momentum conservation, any such rule can be written
\begin{align}
\label{eq:collision}
\nonumber
v_{i}^{*}= & v_{j}+ f(v_{ij}) \, v_{ij}  \\ \nonumber
v_{j}^{*}= & v_{i}-  f(v_{ij}) \, v_{ij}
\end{align}
where $v_{i}$ ($v_{i}^{*}$) and $v_{j}$ ($v_{j}^{*}$) are
the velocities of the particles before (after) the collision,
and $f$ any function of the absolute value of 
the relative velocity $v_{ij}=v_i-v_j$ only, 
and the change in the kinetic energy is 
\begin{equation}
\label{eq:deltaK}
\delta K = -m  f(v_{ij}) \left[1 -  f(v_{ij}) \right] v_{ij}^2
\end{equation}
The family of collision rules we use here are 
ones for which $\delta K < 0$ when $|v_{ij}| < v_0$,
and $\delta K > 0$ when $|v_{ij}| > v_0$, where  
$v_0$ is a constant positive velocity, i.e., 
energy is injected when the particles cross
at a relative velocity of less than $v_0$, and 
is dissipated if it is greater (and conserved when $v=v_0$).
This is a generalisation of a collision rule 
introduced by Brito {\it et al.}\cite{Brito2013} in the 
context of a phenomenological model of an agitated 
granular gas, and for which
\begin{equation}
\label{eq:BRS}
f(v) = q \left(1- \frac{v_0}{|v|}\right)
\end{equation}
where $q$ is a dimensionless constant
(controlling the degree of non-elasticity of the collisions). 
Our results below are for this specific rule, but we 
have explored a variety of quite different functional
forms for $f(v)$ which have the required 
injection/dissipation structure.
The results we obtain are all completely
insensitive to these changes.

We will consider the limit here in which this inelastic collision rule constitutes
a weak perturbation to the system. By this we mean that the time scale
on which it causes non-trivial macroscopic evolution of the system is
long compared to the characteristic time of the mean field
dynamics. It is straightforward to show \cite{Joyce2016} that this 
corresponds to the requirement that $\gamma= qN$
be small, and thus to the quasi-elastic limit $q \rightarrow 0$
\cite{McNamara1993}.

\begin{figure}
\begin{center}
\includegraphics[scale=0.29]{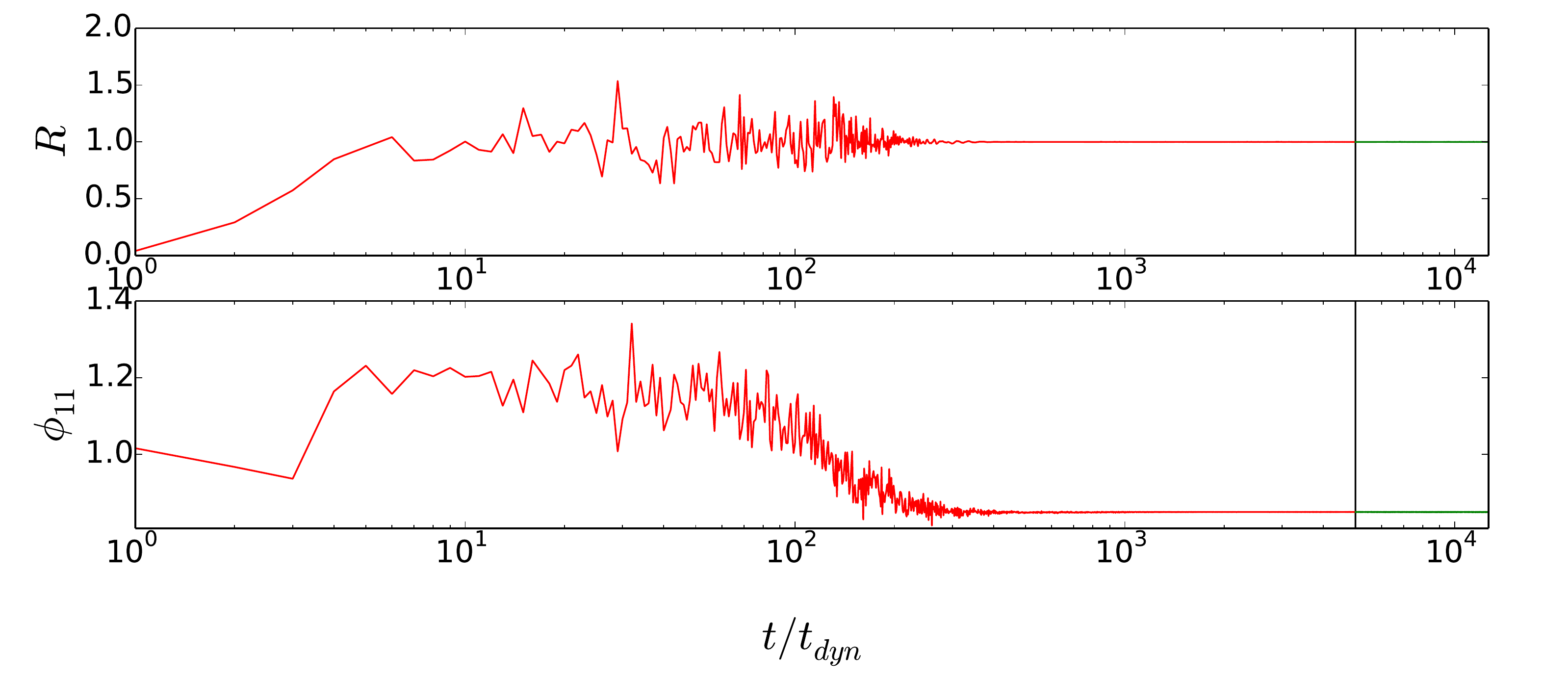}
\end{center}
\caption{Time evolution of the virial ratio $R$ and 
the entanglement parameter $\phi_{11}$ for $N=128$ 
in the sheet model, and $t^*=5000 \tau_{dyn}$ (black vertical line). }
\label{fig:virial}
\end{figure}

We report here results first for the case $\alpha=1$.  Our initial conditions 
are ``rectangular waterbag"  --- particles distributed with random uncorrelated 
positions in an interval $[-L_0/2, L_0/2]$ and random uncorrelated 
velocities in an interval $[-V_0/2, V_0/2]$ --- characterised fully by the 
initial virial ratio $R=2K/U$ (where $K$ and $U$ are kinetic 
and potential energy respectively). 
Simulations are performed using the above collision rule from $t=0$ until 
a time $t^* \gg t_{dyn}/\gamma$ at fixed small $\gamma$ ($\sim 10^{-2}$)
(where $t_{dyn}=2\sqrt{L_0/(gN)}$ is the mean-field  time), and the purely 
Hamiltonian dynamics thereafter. Further details on the algorithm
and numerics can be found in \cite{Joyce2016}. 
For any $R$ we observe the same phases in the evolution. 
Fig.~\ref{fig:virial} shows the temporal evolution of 
two global observables, the virial ratio and the ``entanglement'' parameter, 
$\phi_{11}=\frac{\langle|xv|\rangle}{\langle|x|\rangle\langle|v|\rangle}$,
which is equal to one if the system is in thermodynamic equilibrium \cite{Joyce2010}. 
Firstly, on short time-scales ($t  \ll t_{dyn}/\gamma$), the role of the perturbation is 
negligible and, as the evolution is indistinguishable from that of the Hamiltonian 
system, leads to relaxation 
towards a continuum of virialized states (with properties depending on $R$, see e.g. \cite{Joyce2010}).
Secondly, for $t  \sim t_{dyn}/\gamma$, the system evolves to a macroscopically 
stationary state and then progressively becomes ``ordered'' in the microscopic 
phase space   until it reaches configuration which shows no further
evolution. 
In Fig.~\ref{fig:virial} this non-trivial microscopic evolution
corresponds to the very visible decay of the fluctuations
in the macroscopic observables for $t<t^*$ and no further changes 
for $t>t^*$ when the perturbation is suppressed. 
In the final state $\phi_{11}\simeq 0.85$ which, we note, tells us 
that the system is definitely out of equilibrium.

When the perturbation is switched off, at $t=t^*$, not only is their 
no measurable change in the macroscopic evolution, but the same 
microscopic structure also persists. We have
found this to be true no matter how long we have continued 
to evolve the system, and in particular for times 
very much longer than those observed for 
relaxation to  equilibrium of typical QSS in this system,
of order $(10-100) N t_{dyn}$ \cite{Joyce2010}. 
More precisely, we have observed no measurable
evolution away from the ordered state up to
$t=50000t_{dyn}$ for systems of size up to $N=128$.
Although our results are purely numerical, they strongly suggest
that the state which the system is driven to by perturbation 
is apparently a stable periodic or quasi-periodic solution of
the exact $N$-body dynamics, and thus ergodicity is broken 
and the system will never relax to microcanonical equilibrium.

\begin{figure}[!h]
\begin{center}
\includegraphics[scale=0.21]{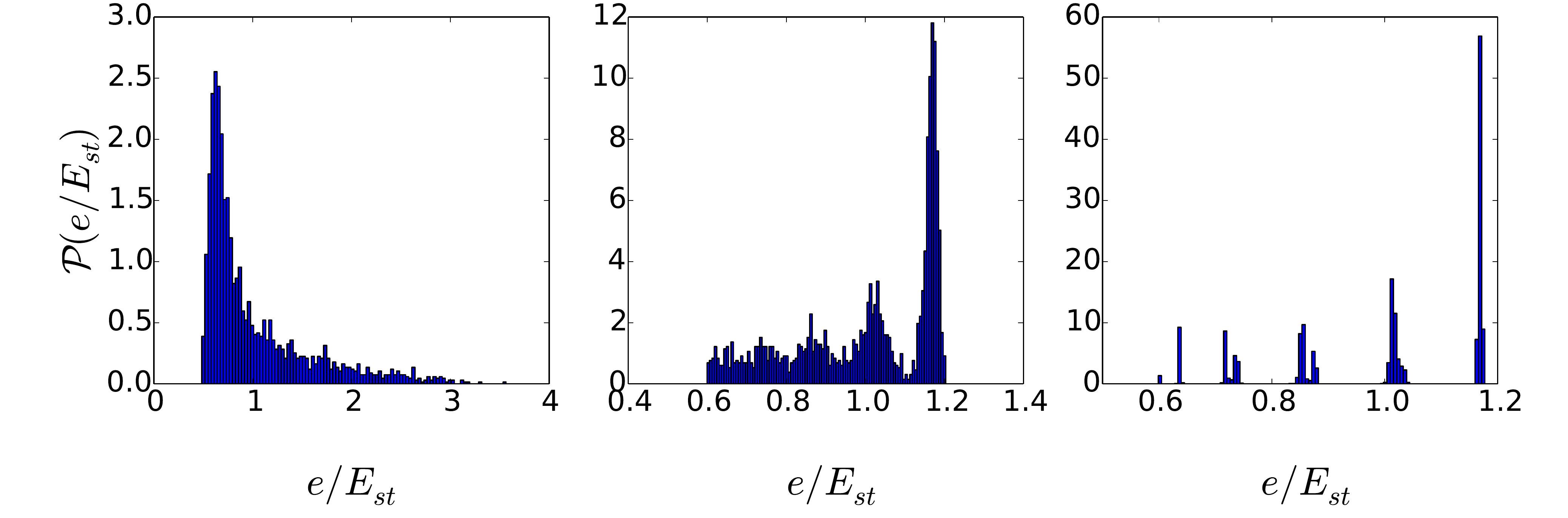}
\end{center}
\caption{PDF of the total energy per particle at different times $t=10,100, 1000 t_{dyn}$ 
for a system of $N=128$.}\label{fig:histo}
\end{figure}

\begin{figure}[!h]
\begin{center}
\includegraphics[scale=0.21]{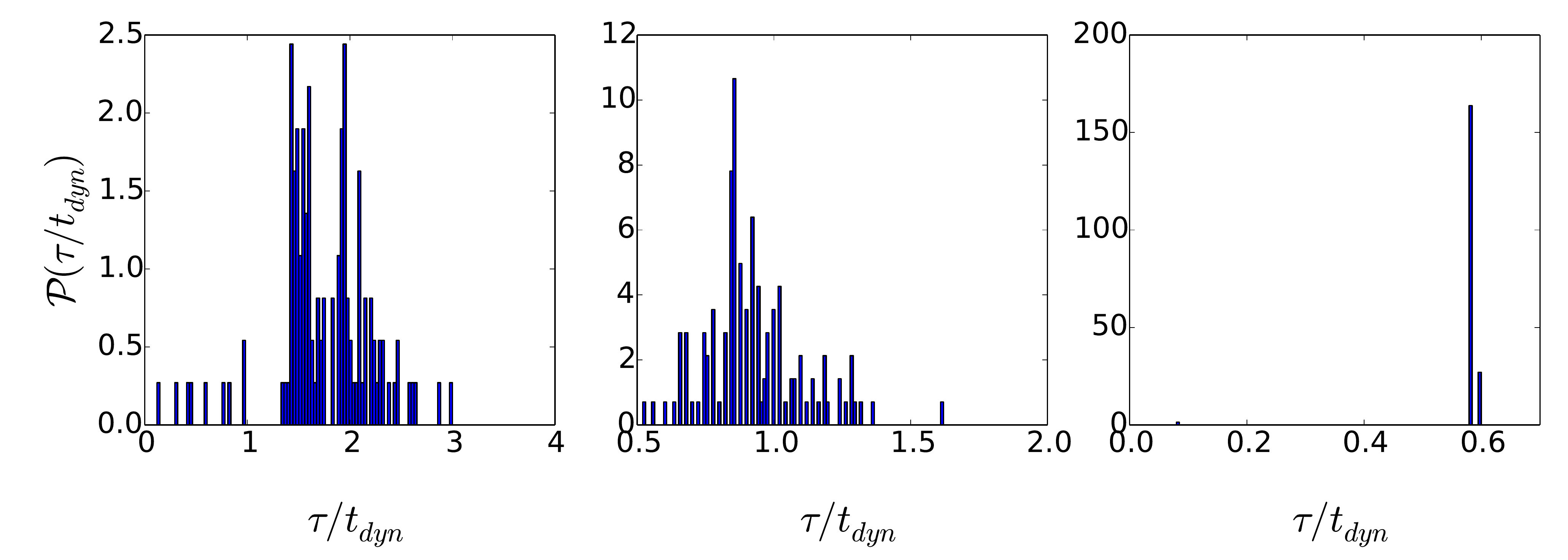}
\end{center}
\caption{PDF of particle ``periods"  at different times for $N=128$.}\label{fig:Frequence}
\end{figure}

To better understand the evolution towards this final state, we have monitored 
the probability distribution function (PDF) of the particle energies 
$e_i=\frac{1}{2}mv^2_i+\sum_{j\neq i}v(x_i-x_j)$.  Fig.\ref{fig:histo} shows,
for a system with $N=128$,  how this PDF  evolves, from a wide distribution at early 
times ($t=10t_{dyn}$) to a very different form ($t=100t_{dyn})$  with several visible 
peaks of amplitude increasing with energy, which then become 
narrower until they are very well localised i.e. almost  $\delta$-function peaks.
The number of peaks is strongly dependent on, and grows with, the system 
size (see Fig.\ref{fig:histo}). 

\begin{figure}[!h]
\begin{center}
\includegraphics[scale=0.3]{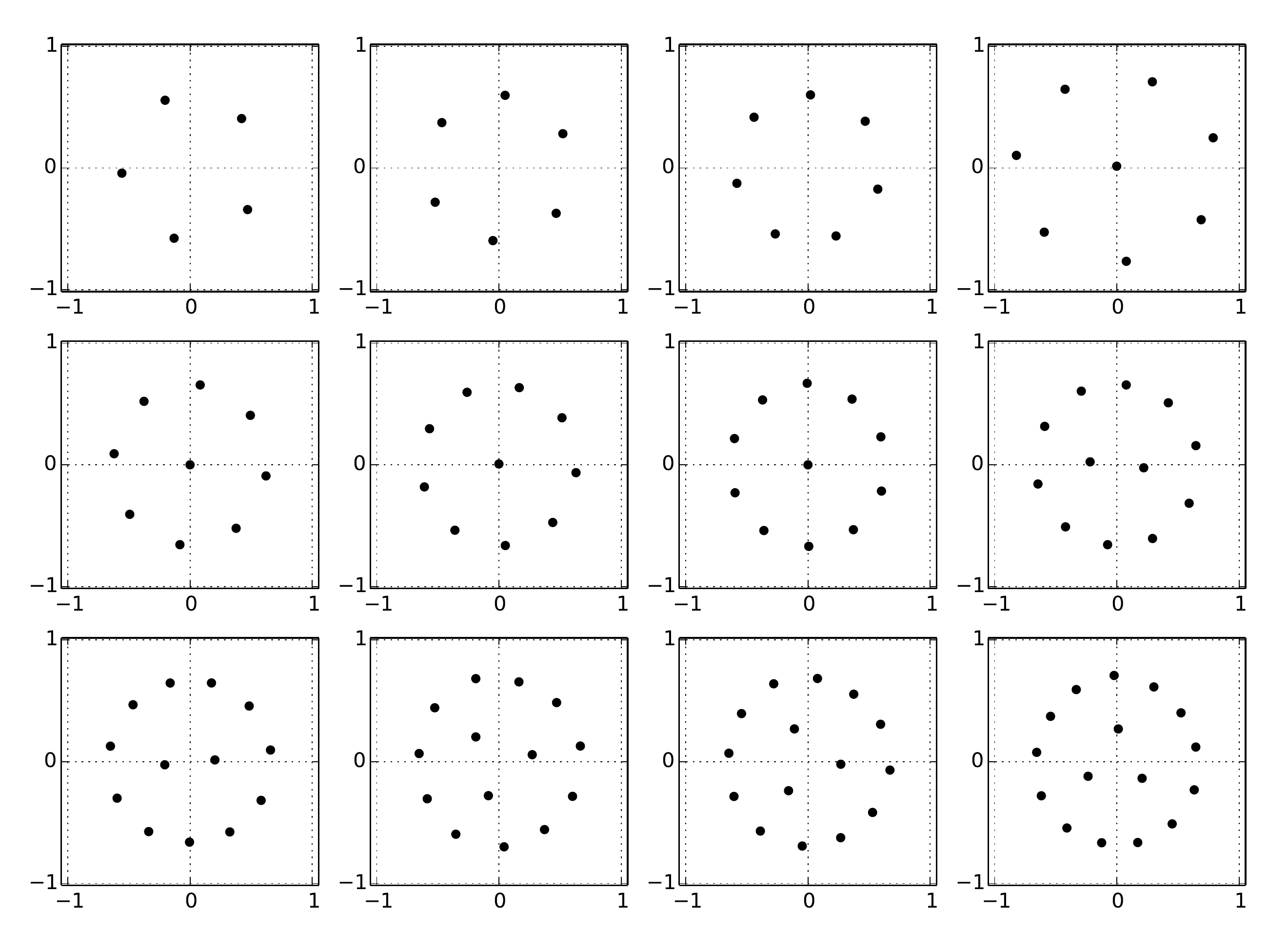}
\end{center}
\caption{Snapshots of the final state after $50000 \tau_{dyn}$ for $N=5,...,16$.}\label{fig:Snapshots}
\end{figure}

\begin{figure}
\begin{center}
\includegraphics[scale=0.32]{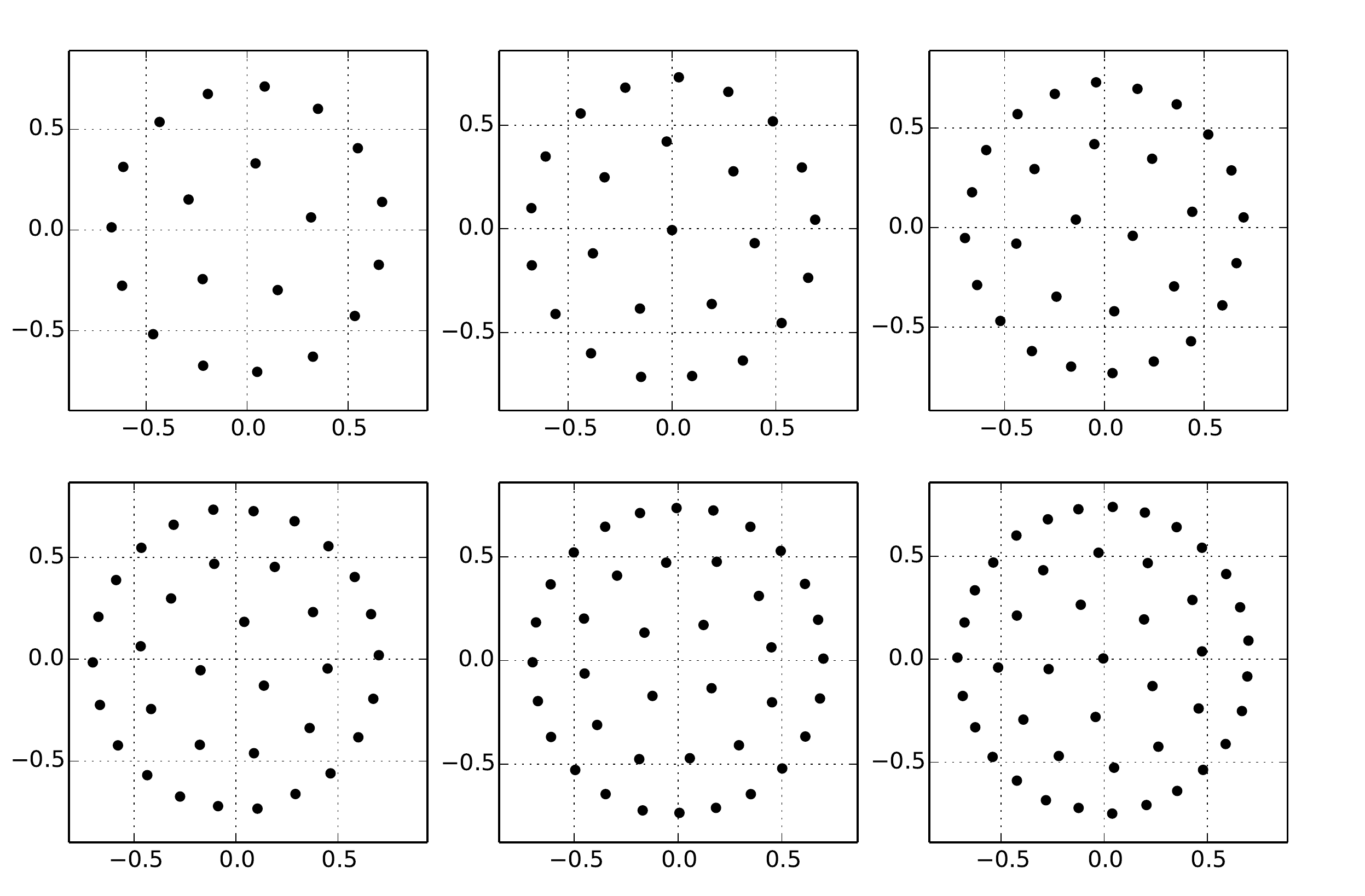}
\end{center}
\caption{Snapshots of the final state after $50000 \tau_{dyn}$ for $N=20,25,30,35,40,45,50$.}\label{fig:Snapshots2}
\end{figure}

The formation of these almost discrete peaks in particle energy corresponds in phase space
to the existence of a ring structure. Further there is also a discrete structure within each ring
(see Figs. \ref{fig:Snapshots}-\ref{fig:Snapshots2}) and a coherence between this 
structure in the different rings. Closer examination of the temporal 
evolution shows that, as one might suspect, the emergence of this ordered structures
reflects a complete coherence of the particle motions.
Figure \ref{fig:Frequence} shows the PDF at different times of the particle ``periods'',
calculated for each particle as the time elapsed since it was last at the same position.    
At short times $t\sim \tau_{dyn}$,  the PDF is broad, but it then evolves on time-scales 
at which the perturbation plays a role, initially changing shape and then converging
to a very narrow peak as the final state is established.  
Again, when the perturbation  is switched off, no measurable evolution of the PDF 
is observed. In the final state therefore all particles appear to move in a completely
coherent periodic (or quasi-periodic) motion.

Given these observations,  the trajectories can be written as a Fourier series
with $x_i=\sum_{n\geq 1}\frac{V_{i,n}T}{2\pi}\cos(2\pi n t/T+\phi_{i,n})$ where $V_{i,n}$ is
the corresponding coefficient of the Fourier expansion of the velocity.  
Figs. \ref{fig:Snapshots} and \ref{fig:Snapshots2}  show  snapshots of the phase
space configuration at $t=50000\tau_{dyn}$,  for various system sizes  
for  $N$  from $5$  to $16$ and for  $N=20,25,30,35, 40, 45,50$.     
These plots are in the dimensionless variables $v/v_0$ and $x/x_0$, where $x_0=v_0T/(2\pi)$
and $T$ is the numerically estimated period. The time evolution of the system is illustrated for two particle sizes ($N=15$ and $N=128$) 
by the  movies provided in the Supplemental Material.
It is clear that, to a good
approximation, the trajectories are circular, which implies that the term $n=1$
in the sum dominates, i.e., the trajectories are very close to those of harmonic
oscillators. 
This phenomenon is only possible if the total force acting on each particle 
is almost independent of the number of particles and depends linearly on the distance 
of the particle to the center of mass of the system. In other words, the system  
reorganises due to the perturbation so that each particle behaves almost like an
independent simple harmonic oscillator.

For systems up to $N=7$, the particles are located on a single circle. For 
$8 \leq N \leq 11$, all but one of the particles are on a single circle while the remaining
one is very close to the origin. For $N=12,13$ the structure in phase space  is now 
composed of two circles of non-zero radii, the smaller with two particles and the larger 
with $N-2$  particles. For $N=14$, a third particle appears on the smaller circle.  
For $N$ varying from $N=20$ to $N=50$, the number of circles increases from $2$ 
ring structure for $N=128,256, 512$ particles for which there 
are $6,8$ and $12$ circles respectively.

Given the regular spacing of the particles on each circle, supposing that there 
are $N_c$ circles, and that the $r$-th circle (where $r=1..N_c$) has $N_r$
particles (and thus $N= \sum_{r=1}^{N_c} N_r$), we can write the particle positions
on a given circle as   $x_i= \frac{V_{}T} {2\pi}\cos(2\pi n t/T+2(i-1)\pi/N_s+\phi_s)$ with 
$i=1..N_s$, and $V_{s,n}/v_0$ is the radius in phase space of
the circle when plotted in the dimensionless variables as in 
Figs. \ref{fig:Snapshots} and \ref{fig:Snapshots2}. 
Note that we have made the approximation, which can be exact
strictly only for the model $\alpha=2$ and not for the sheet
model we are considering,  that $V^s_{i,n}=0$ for $n>1$, i.e., that
the orbits are those of  perfectly decoupled and harmonic oscillators.
The phase $\phi_s$ is the phase of the shell $s$ fixed by the position of 
the particle labeled $i=1$ at $t=0$, and without loss of generality, we can
thus assume $\phi_1=0$.

Given the observed periodicity of the evolution in the stationary state,
we must conclude that the total energy change due to collisions 
over one period is zero. As every particle collides with every other
one twice in one period, this can be written
\begin{equation}\label{eq:deltaE}
\delta E=\sum_{r=1}^{N_c} \sum_{s=1}^{N_c} \sum_{i_r=1}^{N_r}\sum_{j_s=1,j_s\neq i_r}^{N_s} \delta e_{i_rj_s} 
\end{equation}
where $\delta e_{i_rj_s}$ is the energy change due to a collision between a 
particle $i_r$ of the shell $r$ and a particle $j_s$ of the shell $s$. 
As we are working in the quasi-elastic limit, we have,
using Eqs.~(\ref{eq:deltaK}-\ref{eq:BRS}), that 
$\delta e_{i_rj_s}=m\frac{\gamma}{N} [-v^2_{i_r j_s} +  v_0 | v_{i_rj_s} |]$
where $v_{i_rj_s}$ is the relative velocity for a pair of colliding particles.
Since we can assume, in the quasi-elastic limit, that the particles are not 
perturbed from their circular orbits by the collisions which occur when
particles are at the same spatial position, $v_{i_rj_s}$  is simply
given by the (invariant) distance in phase space between the
two particles. We therefore have that
 \begin{equation}\label{eq:vij}
|v_{i_rj_s}|=
\sqrt{V_r^2+V^2_s-2V_r V_s\cos\left(\Phi\right)}, 
\end{equation}
where
$V_r$ and $V_s$ are the amplitudes of the velocities of particles $i_r$ and $j_s$, respectively, and the relative phase 
$\Phi=\frac{2i_r-2}{N_r}-\frac{2j_s-2}{N_l}+\phi_r-\phi_s$.

We first consider systems where all the particles are on a single circle, which we find (numerically)
to be the case for $N\leq 8$. There is then just one free parameter, the dimensionless
radius of the single circle in phase space $R_1=V_1/v_0$ in Fig.\ref{fig:Snapshots}. 
Using Eq.(\ref{eq:vij}), Eq.(\ref{eq:deltaE}) then gives
\begin{equation}\label{eq:diameters}
 R_1=\frac{\sum_{i=1}^{N-1} \sin\left(i\frac{\pi}{N}\right)}{2\sum_{i=1}^{N-1} \sin^2\left(i\frac{\pi}{N}\right)  }
\end{equation}
In Table~\ref{tab:radius}, we compare simulation results and theoretical predictions, Eq.(\ref{eq:diameters}).
For these small system sizes ($N=5,6,7$) 
the uncertainty of the radii reported in Table~\ref{tab:radius} reflects the deviations 
from the assumed circular trajectories which decrease as $N$ grows.

For $N \geq 8$, our simulations starting from waterbag initial conditions always give rise to
a final state with a single particle very close to the origin and the others very close to
periodic motion on a single circle. Assuming the configurations to be given by one
with one particle assumed to stay at rest at the origin and the other $N-1$ particles 
on the circle, there is again a single free parameter $R_1$ which is given by a simple 
expression derived from Eq. (\ref{eq:vij}) taking $V_2=0$ and $V_1=v_0 R_1$. The comparison of the 
numerical and theoretical results is shown again in Table~\ref{tab:radius}. 
For  $N=9,10,11$, we obtain the same stable configuration and have obtained $R_1$ using the same reasoning.
For $N \geq 12,13$, our simulations, as we have noted give final configuration with at least two
circles both with more than one particle. There is then more than one free parameter and 
the condition $\delta E=0$ can thus not determine the state uniquely. For two such circles  
we find, however, that we can obtain good predictions from this condition alone with some
simple assumptions. In general there are in this case, given the observed number of 
particles on the two circles, three free parameters: the two  dimensionless radii of the 
circles $R_1=V_1/v_0$ and $R_2=V_2/v_0$, and the phase $\phi_2$. 
In practice the condition is very weakly dependent on $\phi_2$, and so we set
$\phi_2=0$. We then obtain the condition in the form of an implicit equation $f(R_1,R_2)=0$. 
We find that choosing the unique solution corresponding to the largest value of $R_1$, 
which corresponds to the largest value of the energy, we observe good agreement with 
simulation results. For the case of two circles plus a single particle at rest, an analogous
calculation can be performed and gives again (see Table \ref{tab:radius}) satisfactory
results for case $N=25$.

Numerically we find that in all cases the radius $R_1$ of the biggest circle always increases when
a circle is added, and thus we expect that the limit $N \rightarrow \infty $ of Eq.~(\ref{eq:diameters}),
which gives $R_1=2/\pi$, should be a strict lower bound on its value. For larger systems 
(up to $N=512$) we observe also that the number of particles on the outer circle 
$N_1\simeq N/2$, and the value of $R_1$ in the configuration with the other $N-N_1$ particles 
at the origin, should thus give an upper bound on it. It is simple to show that 
this is $R_1=1/ \pi +1/2$. Using the same reasoning, assuming
all the particles on one circle, one obtains a lower bound on the parameter,  
$\phi_{11}=\pi/4=0.785..$, which is smaller than and comparable to the 
simulation results ($\phi_{11}=0.85$). We note that our observation of these apparently stable 
states for $N\geq 11$ appears to invalidate the analysis of \cite{Reidl1993}, who concluded 
that this system should be strictly ergodic in this case.

\begin{table}[t]
\caption{Measured and predicted radii 
}
\begin{ruledtabular}
\begin{tabular}{lcr}
\textrm{$N$}&
\textrm{Simulation}&
\textrm{Predictions}
\\
\colrule
  $5$ & $0.59 \pm 0.02$ & $0.615$\\
  $6$ & $0.60 \pm 0.02$& $0.622$ \\
  $7$ & $0.61 \pm 0.02 $ & $0.6258$\\ 
  $8$ & $0.785\pm 0.03$&$0.6726$ (2 circles)\\
  $9$ & $0.602\pm0.03$&$0.6697$ (2 circles)\\
  $10$ &  $0.666$  &$ 0.667$(2 circles)\\
  $11$ &  $0.660$  & $0.665$   (2 circles) \\
  $12$ &   $0.23,0.66$  &$0.25,0.702$  (2 circles)   \\
  $13$ &   $0.21,0.60$  &$0.25,0.697$ (2 circles)\\
  $20$ & $0.32,0.717$ & $0.31,0.74$ (2 circles)\\
  $25$ & $0,0.37, 0.74$ & $0,0.32,0.79,$ (3 circles)
  \\
\colrule
\textrm{$N$}&
\textrm{Simulation}
&
\textrm{lower and upper bounds}
\\
\colrule
  $30$ & $0.738$ & $0.637,0.818$ \\
  $35$ & $0.700$ &  $0.637,0.818$  \\
  $40$ & $0.705$ & $0.637,0.818$\\
  $45$ & $0.705$ & $0.637,0.818$\\
  $50$ & $0.715$ & $0.637,0.818$
\end{tabular}
\end{ruledtabular}\label{tab:radius}
\end{table}
As detailed in the Supplemental Material, we have studied numerically
a broad range of values of $\alpha$, introducing for the case $\alpha <1$
a ``smoothing'' of the divergence of the force at $r=0$ at a scale $\epsilon$.
We have also studied the paradigmatic HMF model\cite{Campa2009}. In the range 
$0<\alpha \leq 2$ we observe always a rapid emergence of 
ordered structures in phase space like those in the case $\alpha=1$,
with no dependence on the scale $\epsilon$ provided it is sufficiently
small. For $-1 <\alpha < 0$ we again observe such states, but only
provided  $\epsilon$ is sufficiently large. This  result  is not unexpected as  
the interparticle force becomes integrable at large distances for $\alpha <0$. 
This has the consequence that,
unless a sufficiently large smoothing of the force is applied, the rapidly
fluctuating contributions to the force from nearby particles will dominate 
over the bulk contribution, destroying the characteristic long-range
behaviour of the dynamics \cite{Gabrielli2010a}. The states obtained in 
the different models differ in detail, showing, in particular, different  
values of  $N$ at which the number of circles change. For $\alpha>2$ 
we find, on the other hand, that we observe the states only at low $N$, 
and only in a single ring structure. For the HMF we observe the states,
but only provided $v_0$ is sufficiently small that the stationary state
is highly magnetised. In both these cases it appears that the states do
not appear because the particles cannot in general self-organize 
to produce an approximately harmonic potential.

It seems reasonable to posit that the existence of these states may be a
universal feature of a very broad range of long-range systems, also in higher 
dimensions, and further, that it may be possible to attain them efficiently 
by applying non-Hamiltonian  perturbations of the kind we have considered. These questions, 
along with many other ones concerning the origin and nature of these intriguing states, 
and an exploration of the rich and varied states which may be obtained with 
a broader class of inelastic collision rules will  be addressed  in forthcoming studies.  

We acknowledge D. Benhaiem and F. Sicard for assistance with numerical codes, and 
B. Bacq-Labreuil for performing some simulations. We also thank warmly A. Gabrielli  for 
very useful discussions. P.V. acknowledges H. Touchette for fruitful discussions and the 
kind hospitality of NITheP, Stellenbosch where a part of this work was performed.

%

\end{document}